\pdfoutput=1

\documentclass[11pt,twoside,a4paper,cmspaper,final,collab]{cms-tdr}
\def\svnVersion{74b16f8-D}\def\svnDate{2019/09/26}\def\cmsCernNoTag{CERN-EP-2018-283}\def\cmsCernDate{\today}\def\cmsMessage{"Published in Physics Letters B as \href{http://dx.doi.org/10.1016/j.physletb.2019.135048}{\doi{10.1016/j.physletb.2019.135048}.}"}

\begin{document}\cmsNoteHeader{HIN-17-007}

\hyphenation{had-ron-i-za-tion}
\hyphenation{cal-or-i-me-ter}
\hyphenation{de-vices}
\RCS$HeadURL$
\RCS$Id$

\newlength\cmsFigWidth
\ifthenelse{\boolean{cms@external}}{\setlength\cmsFigWidth{0.85\columnwidth}}{\setlength\cmsFigWidth{0.4\textwidth}}
\ifthenelse{\boolean{cms@external}}{\providecommand{\cmsLeft}{top\xspace}}{\providecommand{\cmsLeft}{left\xspace}}
\ifthenelse{\boolean{cms@external}}{\providecommand{\cmsRight}{bottom\xspace}}{\providecommand{\cmsRight}{right\xspace}}
\ifthenelse{\boolean{cms@external}}{\providecommand{\cmsTable}[1]{#1}}{\providecommand{\cmsTable}[1]{\resizebox{\textwidth}{!}{#1}}}

\newcommand {\pPb}{\ensuremath{\Pp\text{Pb}}\xspace}
\newcommand{\WToMuNu}{\ensuremath{\PW\to\PGm\PGnGm}\xspace}
\newcommand{\WToMuNuPl}{\ensuremath{\PWp\to\PGmp\PGnGm}\xspace}
\newcommand{\WToMuNuMi}{\ensuremath{\PWm\to\PGmm\PAGnGm}\xspace}
\newcommand{\WToMuNupm}{\ensuremath{\Wpm\to\PGmpm\PGnGm}\xspace}
\newcommand{\DY}{\ensuremath{\PZ/\Pgg^{*}}\xspace}
\newcommand{\DYToMuMu}{\ensuremath{\DY\to\MM}\xspace}
\newcommand{\DYToTauTau}{\ensuremath{\DY\to\TT}\xspace}
\newcommand{\ZToMuMu}{\ensuremath{\PZ\to\MM}\xspace}
\newcommand{\WToTauNu}{\ensuremath{\PW\to\PGt\PGnGt}\xspace}
\newcommand{\POWHEGBOX}{{\textsc{powheg}}\xspace\xspace{\textsc{box}}\xspace}
\newcommand{\EPOSLHC} {{\textsc{epos~lhc}}\xspace}
\newcommand{\etaLab}{\ensuremath{\eta_{\text{lab}}}\xspace}
\newcommand{\etaCM}{\ensuremath{\eta_{\text{CM}}}\xspace}
\newcommand{\etaMuLab}{\ensuremath{\etaLab^{\PGm}}\xspace}
\newcommand{\etaMuCM}{\ensuremath{\etaCM^{\PGm}}\xspace}
\newcommand{\IsoMu}{\ensuremath{I_{\PGm}}\xspace}
\newcommand{\ptMu}{\ensuremath{\pt^{\PGm}}\xspace}
\newcommand{\utvec}{\ensuremath{{\vec u}_{\mathrm{T}}}\xspace}
\newcommand{\RFB}{\ensuremath{R_\mathrm{FB}}\xspace}
\newcommand{\RFBChgMu}{\ensuremath{N_{\PGm}^{\pm}(+\etaMuCM)/N_{\PGm}^{\pm}(-\etaMuCM)}\xspace}
\newcommand{\RFBAllMu}{\ensuremath{N_{\PGm}(+\etaMuCM)/N_{\PGm}(-\etaMuCM)}\xspace}
\newcommand{\ChgAsym}{\ensuremath{(N_{\PGm}^{+} - N_{\PGm}^{-})/(N_{\PGm}^{+} + N_{\PGm}^{-})}\xspace}

\cmsNoteHeader{HIN-17-007}

\title{Observation of nuclear modifications in {\Wpm} boson production in \pPb collisions at $\sqrtsNN = 8.16\TeV$}

\date{\today}

\abstract{
The production of {\Wpm} bosons is studied in proton-lead (\pPb) collisions at a nucleon-nucleon centre-of-mass energy of $\sqrtsNN = 8.16\TeV$. Measurements are performed in the \WToMuNupm channel using a data sample corresponding to an integrated luminosity of $173.4 \pm 6.1\nbinv$, collected by the CMS Collaboration at the LHC. The number of positively and negatively charged {\PW} bosons is determined separately in the muon pseudorapidity region in the laboratory frame $\abs{\etaMuLab} < 2.4$ and transverse momentum $\ptMu > 25\GeVc$. The {\Wpm} boson differential cross sections, muon charge asymmetry, and the ratios of {\Wpm} boson yields for the proton-going over the Pb-going beam directions are reported as a function of the muon pseudorapidity in the nucleon-nucleon centre-of-mass frame. The measurements are compared to the predictions from theoretical calculations based on parton distribution functions (PDFs) at next-to-leading-order. The results favour PDF calculations that include nuclear modifications and provide constraints on the nuclear PDF global fits.
}

\hypersetup{
pdfauthor={CMS Collaboration},
pdftitle={Observation of nuclear modifications in W boson production in pPb collisions at sqrt(s[NN]) = 8.16 TeV},
pdfsubject={CMS},
pdfkeywords={CMS, physics, heavy ions, electroweak, W boson, pPb, nPDF, 8.16 TeV}
}

\maketitle

\section{Introduction}

The production of electroweak (EW) gauge bosons is considered to be a powerful probe of the parton distribution functions (PDFs) of the proton~\cite{Butterworth:2015oua}. Most recent proton PDF sets include {\PW} and {\PZ} boson production data from the Tevatron and the CERN Large Hadron Collider (LHC) in their global fit analyses~\cite{Dulat:2015mca,Ball:2014uwa,Harland-Lang:2014zoa}. Similarly, the measurements of EW boson production in proton-nucleus and nucleus-nucleus collisions, available for the first time at centre-of-mass energies of the {\TeVns} scale, provide constraints on nuclear modifications of the PDFs~\cite{Vogt:2000hp,Zhang:2002yz,Paukkunen:2010qg,Kusina:2016fxy}. The presence of a nuclear environment modifies the parton densities in the nucleus as compared to those in a free nucleon. The nuclear PDFs (nPDFs) are expected to be enhanced for partons carrying a momentum fraction in the range $5\times10^{-2} \lesssim x \lesssim 10^{-1}$ in the so-called \textit{antishadowing} region, and suppressed for $x\lesssim10^{-2}$ in the \textit{shadowing} region~\cite{Armesto:2006ph}, with the modifications depending on the scale $Q^2$. Because of the limited amount and type of experimental data sets available for nuclear collisions, the determination of the nuclear parton densities is less precise than for the free-proton case. As a consequence, the nPDF uncertainties are one of the main limitations of the precision of quantum chromodynamics (QCD) calculations describing hard-scattering processes in nuclear collisions at high energies~\cite{Paukkunen:2010qg}.

Since {\PW} bosons are predominantly produced via \qqbar annihilation through $\cPqu\cPaqd\to\PWp$ and $\cPqd\cPaqu\to\PWm$ processes, {\PW} boson production can be used to probe the light quark PDFs, both for the proton and nuclei. In addition, the asymmetries of the separate yields of {\PWp} and {\PWm} bosons are known to be sensitive probes of the down-to-up quark PDF ratio~\cite{Aaltonen:2009ta,Chatrchyan:2013mza,Chatrchyan:2012xt}. Consequently, their measurement may allow for the flavour decomposition of {\cPqu} and {\cPqd} quark distributions in nuclei~\cite{Khachatryan:2015hha}. Among the possible decay channels of the {\PW} boson, the leptonic decays are less affected by background processes than hadronic decays. Another advantage of the leptonic decays is that any possible effect due to the QCD medium produced in nuclear collisions should be negligible, since leptons are not subject to medium-induced energy loss through the strong interaction~\cite{ConesadelValle:2007sw,ConesadelValle:2009vp}.

Studies of the {\PW} and {\PZ} boson production in PbPb collisions at a nucleon-nucleon centre-of-mass energy of $\sqrtsNN = 2.76\TeV$, performed by the ATLAS~\cite{Aad:2010aa,Aad:2012ew,Aad:2014bha} and CMS~\cite{Chatrchyan:2011ua,Chatrchyan:2014csa,Chatrchyan:2012nt} Collaborations, have shown that the {\PW} and {\PZ} boson cross sections are consistent with no modification by the nuclear medium formed in these collisions. In \pPb collisions, measurements of {\PW} production at $\sqrtsNN = 5.02\TeV$ have been performed by ALICE~\cite{Alice:2016wka} and CMS~\cite{Khachatryan:2015hha}. The comparison with next-to-leading-order (NLO) perturbative QCD predictions favours the calculations that include nPDF effects. A similar observation is made from the analysis of the {\PZ} boson production in \pPb collisions at the same energy~\cite{Khachatryan:2015pzs,Aad:2015gta}. These EW boson measurements have been used for the first time in a global fit analysis of nPDF sets (EPPS16~\cite{Eskola:2016oht}). Nevertheless, a modest enhancement of the {\PWm} boson production cross section in the most backward region (Pb-going direction) showed some difference with theoretical calculations (with and without nPDF effects), possibly pointing to different nuclear modifications of the up and down quark PDFs~\cite{Khachatryan:2015hha}. More precise measurements are thus needed in order to clarify the origin of this discrepancy.

This letter reports the results of measurements of {\Wpm} boson production in \pPb collisions at $\sqrtsNN = 8.16\TeV$. The measurements are performed in the \WToMuNupm decay channel using \pPb data recorded with the CMS detector in 2016, corresponding to a total integrated luminosity of $173.4 \pm 6.1\nbinv$. This data set is roughly five times larger than the one available for the previous measurement at $\sqrtsNN = 5.02\TeV$ \cite{Khachatryan:2015hha,CMS-PAS-LUM-13-002,CMS-PAS-LUM-17-002}. The {\Wpm} boson differential cross sections are presented as functions of the muon pseudorapidity in the nucleon-nucleon centre-of-mass (CM) frame, \etaCM. In order to fully exploit the information provided by the data, two additional sets of observables are also measured as functions of \etaCM: the muon charge asymmetry and the muon forward-backward ratios (\RFB). The measurement of asymmetries has a couple of advantages as compared to that of the cross sections. First, asymmetries are more sensitive to modifications of the quark PDFs~\cite{Paukkunen:2010qg}. Second, uncertainties in the integrated luminosity and the theoretical scale dependence cancel in the measurement of these asymmetries. The results of the {\Wpm} boson differential cross sections and asymmetries are compared to perturbative QCD calculations based on NLO PDFs with and without nuclear modifications. The theoretical predictions for free protons are obtained using the CT14~\cite{Dulat:2015mca} proton PDF set, while those including nuclear effects are derived using two different nPDF sets for lead ions: nCTEQ15~\cite{Kovarik:2015cma} and EPPS16~\cite{Eskola:2016oht}.

\section{Experimental methods}

\subsection{Data-taking conditions and the CMS detector}

During the data-taking period, the directions of the proton and lead beams were swapped after an integrated luminosity of 62.6\nbinv was collected. The beam energies were 6.5\TeV for the protons and 2.56\TeV per nucleon for the lead nuclei. By convention, the proton-(Pb-)going side defines the positive (negative) $\eta$ region, labelled as the forward (backward) direction. Because of the asymmetric collision system, massless particles produced in the nucleon-nucleon centre-of-mass frame at an \etaCM are reconstructed at $\etaLab = \etaCM + 0.465$ in the laboratory frame. The {\Wpm} boson measurements presented in this letter are expressed in terms of the muon pseudorapidity in the CM, \etaMuCM.

The central feature of the CMS apparatus is a superconducting solenoid of 6\unit{m} internal diameter, that provides a magnetic field of 3.8\unit{T}. Within the solenoid volume are a silicon pixel and strip tracker, a lead tungstate crystal electromagnetic calorimeter (ECAL), and a brass and scintillator hadron calorimeter (HCAL), each composed of a barrel and two endcap sections. Forward calorimeters extend the $\eta$ coverage provided by the barrel and endcap detectors. The hadron forward (HF) calorimeter uses steel as an absorber and quartz fibres as the sensitive material. The two halves of the HF are located 11.2\unit{m} from the interaction region, one on each end, and together they provide coverage in the range $3.0 < \abs{\eta} < 5.2$. They also serve as luminosity monitors. Muons are measured in the range $\abs{\eta} < 2.4$, with detection planes made using three technologies: drift tubes, cathode strip chambers, and resistive plate chambers. A more detailed description of the CMS detector, together with a definition of the coordinate system used and the relevant kinematic variables, can be found in Ref.~\cite{Chatrchyan:2008zzk}.

The particle-flow (PF) algorithm~\cite{Sirunyan:2017ulk} aims to reconstruct and identify each individual particle in an event, with an optimised combination of information from the various elements of the CMS detector. The energy of photons is obtained from the ECAL measurement. The energy of electrons is determined from a combination of the electron momentum at the primary interaction vertex as determined by the tracker, the energy of the corresponding ECAL cluster, and the energy sum of all bremsstrahlung photons spatially compatible with originating from the electron track. The charge and momentum of muons is obtained from the curvature of the corresponding track. The energy of charged hadrons is determined from a combination of their momentum measured in the tracker (assuming the charged-pion mass) and the matching ECAL and HCAL energy deposits, corrected for zero-suppression effects and for the response function of the calorimeters to hadronic showers. Finally, the energy of neutral hadrons is obtained from the corresponding corrected ECAL and HCAL energies.

\subsection{Event selection and muon reconstruction}

Collision events are required to have at least one interaction vertex reconstructed using two or more tracks within a distance from the nominal collision point of 25\cm along the beam axis and 2\cm along its transverse plane. The contamination from background events not originating from inelastic hadronic collisions is further suppressed by requiring at least one tower on each side of the HF calorimeter with a total energy larger than 3\GeV. The loss of events with {\Wpm} bosons candidates due to this \pPb collision event selection has been determined to be less than 0.2\%.

The main signature of the \WToMuNupm process is the presence of an isolated high-\pt muon. Events of interest for offline analysis are selected using a trigger algorithm~\cite{Khachatryan:2016bia} that requires the presence of at least one muon candidate of $\pt > 12\GeVc$. Moreover, to enhance the signal purity~\cite{Khachatryan:2015hha,Chatrchyan:2013mza}, the fiducial region of the analysis has been restricted to muons of $\pt > 25\GeVc$ with $\abs{\etaMuLab} < 2.4$.  The muon candidates are reconstructed in CMS with an algorithm that combines the information from the muon detectors and the tracker~\cite{Sirunyan:2018fpa}. Muons are selected by applying the standard tight selection criteria described in Ref.~\cite{Sirunyan:2018fpa}.  Further, muons are required to be isolated from nearby hadronic activity to reduce the jet background. The muon isolation parameter (\IsoMu) is defined as the \pt sum of all PF-reconstructed photons, charged and neutral hadrons, in a cone of radius $R = \sqrt{\smash[b]{(\Delta\eta)^2+(\Delta\phi)^2}} = 0.3$ around the muon candidate, where $\Delta\eta$ and $\Delta\phi$ are the pseudorapidity and azimuthal (in radians) distances to the muon. A muon is considered isolated if \IsoMu is less than 15\% of the muon \pt.

Background processes yielding high-\pt muons can be classified as reducible or irreducible. The reducible background includes muon decays that can be tagged and removed from the signal. These events are mainly composed of \MM pairs from Drell--Yan events (\DY), and high-\pt muons from jets produced via the strong interaction, referred to as QCD multijet events.  To further suppress the former processes, events containing at least two isolated oppositely charged muons, each with $\ptMu > 15\GeVc$, are removed.  The irreducible background sources comprise muon decays that pass the analysis selection criteria and therefore cannot be tagged event-by-event, including \DYToTauTau, \WToTauNu, and \ttbar production.
All backgrounds are estimated using Monte Carlo (MC) simulations, except QCD multijet, which is modelled with a data-driven technique described below.

\subsection{Signal yield determination}

Leptonic decays of {\PW} bosons include neutrinos, which are not detected in CMS. Their presence is inferred from the overall momentum imbalance in the transverse plane, known as the missing transverse momentum \ptvecmiss; its magnitude (\ptmiss) is defined as the magnitude of the sum of the negative \pt vectors of all reconstructed PF objects in an event. In this analysis, the \ptmiss distribution is used to extract the signal yields in 24 muon \etaMuCM bins, each 0.2 units wide, except for four in the most backward ($-2.86 < \etaMuCM < -2.60$, $-2.20 < \etaMuCM < -1.93$, $-1.93 < \etaMuCM < -1.80$), and forward ($1.80 < \etaMuCM < 1.93$) regions, because of the detector geometry and the unbalanced beam energies.

{\tolerance=800
The \ptmiss distributions of the signal and EW backgrounds are described using templates from MC simulations. The MC samples were generated using the NLO generator \POWHEG v2 \cite{Frixione:2007vw,Nason:2004rx,Alioli:2010xd}. To include EW corrections, the \POWHEGBOX packages W\_ew-BMNNP~\cite{Barze:2012tt} and Z\_ew-BMNNPV~\cite{Barze:2013fru} are used to generate the $\Pp\Pp\to\Wpm\to\Pl^{\pm}\PGnl$ and $\Pp\Pp\to\DY\to\Pl^{+}\Pl^{-}$ processes, respectively. Events from the $\Pp\Pp\to\ttbar$ process are generated using the \POWHEGBOX package hvq \cite{Frixione:2007nw}, which is a heavy flavour quark generator. The simulation of \pPb collisions is performed using the CT14~\cite{Dulat:2015mca} PDF set corrected with the EPPS16 nuclear modification factors, defined as the ratios of the bound proton PDF to that of a free proton, derived for Pb ions~\cite{Eskola:2016oht}. The parton densities of protons and neutrons are scaled according to the mass and atomic number of the lead isotopes.
\par}

The parton showering is performed by hadronising the events using \PYTHIA 8.212~\cite{Sjostrand:2014zea} with the CUETP8M1 underlying-event (UE) tune~\cite{Khachatryan:2015pea,Chatrchyan:2013gfi}. To consider a more realistic distribution of the underlying environment present in \pPb collisions, the \POWHEG samples are embedded in simulated \pPb events generated by \EPOSLHC (v3400)~\cite{Pierog:2013ria}, taking into account the \pPb boost. The \EPOSLHC simulation is tuned to reproduce the global event properties of the \pPb data such as the $\eta$ distributions of charged hadrons~\cite{Sirunyan:2017vpr}. The embedding of the signal and \pPb UEs is performed by requiring the same generated interaction point when simulating the detector hits. The trigger decisions are emulated and the embedded events are reconstructed with the same algorithms as used for data. The detector response is simulated with \GEANTfour~\cite{Agostinelli:2002hh}.

The agreement between the EW simulations and the data is improved by weighting the EW boson \pt distribution using a \pt-dependent function derived from the ratio of the {\PZ} boson \pt distributions in \ZToMuMu events in data and simulation. Furthermore, the \pPb event activity is reweighed by matching the simulated total energy distribution reconstructed on both sides of the HF calorimeters to the one observed in data in a \DYToMuMu sample.

The shape of the QCD multijet background is modelled with a functional form described by a modified Rayleigh distribution~\cite{Chatrchyan:2014mua} defined as:
\begin{linenomath}
\ifthenelse{\boolean{cms@external}}
{ 
\begin{multline*}
f\left( \ptmiss \right) = \ptmiss \\
\exp\left[ \left. - \left( \ptmiss \right)^{2} \right/ 2\left( \sigma_{0} + \sigma_{1} \ptmiss + \sigma_{2}\left( \ptmiss \right)^{2} \right)^{2} \right],
\end{multline*}
} 
{ 
\begin{equation*}
f\left( \ptmiss \right) = \ptmiss \exp\left[ \left. - \left( \ptmiss \right)^{2} \right/ 2\left( \sigma_{0} + \sigma_{1} \ptmiss + \sigma_{2}\left( \ptmiss \right)^{2} \right)^{2} \right],
\end{equation*}
}
\end{linenomath}
where $\sigma_{0}$, $\sigma_{1}$, and $\sigma_{2}$ are free parameters to be determined. It is found to reproduce well the \ptmiss shape of data events containing nonisolated muons, with $\chi^2$ values divided by the number of degrees of freedom (dof) close to one. The QCD shape is extracted by fitting the data in five relative muon isolation (\IsoMu/\ptMu) bins with boundaries ranging from 0.4 to 0.9. The $\sigma_{0}$, $\sigma_{1}$, and $\sigma_{2}$ parameters extracted from the fits are found to linearly depend on the relative muon isolation and are extrapolated to the isolated muon signal region.

Because of momentum conservation, the production of {\PZ} and {\PW} bosons is balanced by a hadronic recoil composed of jets and particles from the \pPb underlying activity. The distribution of the hadronic recoil significantly contributes to the \ptmiss resolution. Because of the similarity of the production processes of the {\PZ} and {\PW} bosons, and their similar masses, we assume that the recoil distributions are the same for both species. Therefore, the correction of the simulated recoil distribution is derived in a control region of \ZToMuMu events using a hadronic recoil technique~\cite{Chatrchyan:2011tn,CMS-PAS-JME-16-004}. The hadronic recoil of \ZToMuMu events, \utvec, is defined as the vector \pt sum of all PF candidates, excluding the decay products of the {\PZ} boson. The distributions of the hadronic recoil components that are parallel and perpendicular to the {\PZ} boson transverse momentum $\ptvec^{\PZ}$ are fitted in simulation and data using a weighted sum of two Gaussian functions. The mean and resolution values extracted from the recoil fits are used to scale the simulated hadronic recoil distributions to match the performance measured in data. The corrected \ptmiss distribution is then derived in the EW MC samples as the vector sum of the corrected hadronic recoil $\utvec^{\text{corr}}$ and the \ptvec of the reconstructed muons from the decay of {\PZ} and {\PW} bosons.

\begin{figure}[htbp]
 \centering
 \includegraphics[width=1.2\cmsFigWidth]{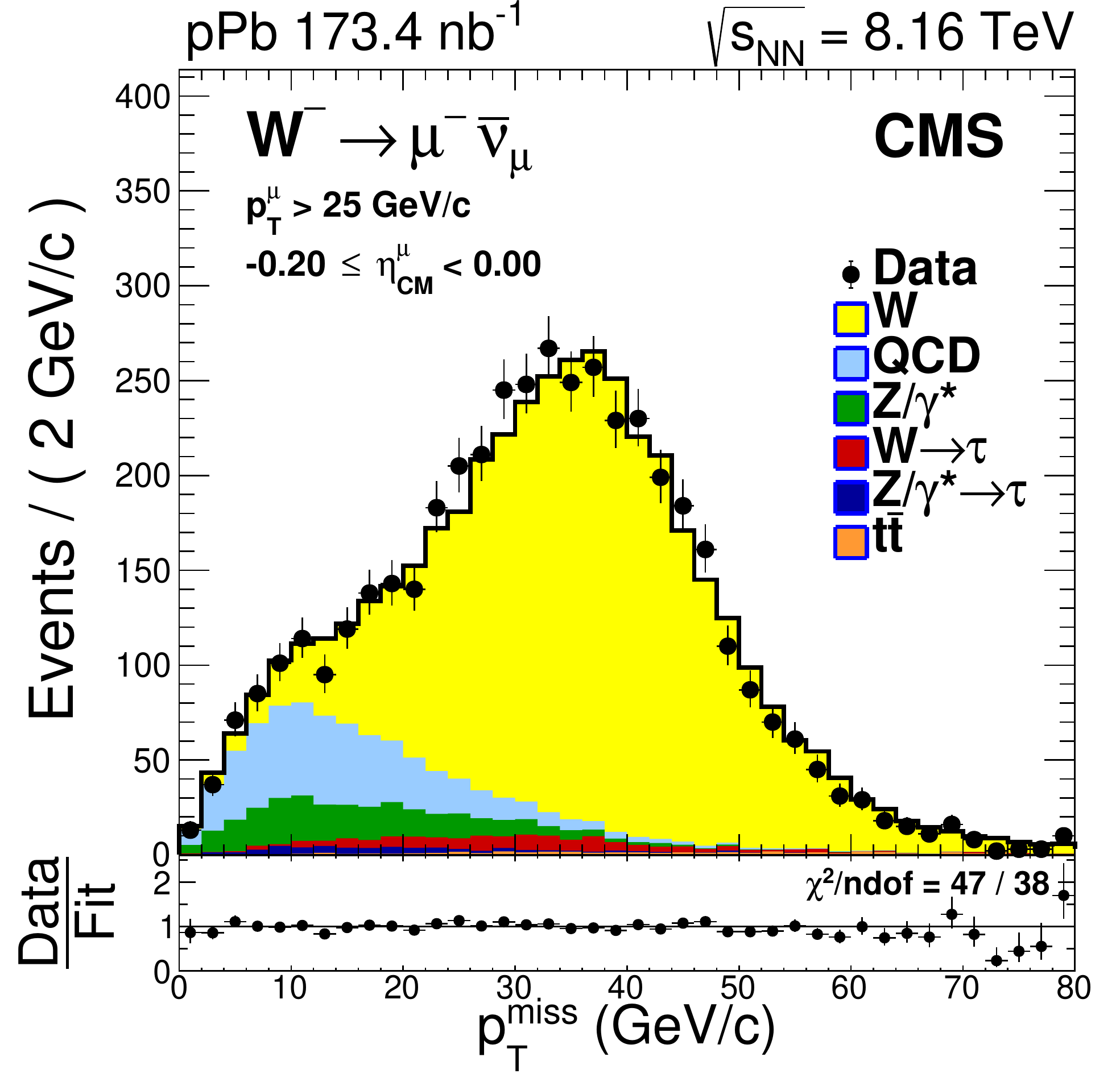}
 \includegraphics[width=1.2\cmsFigWidth]{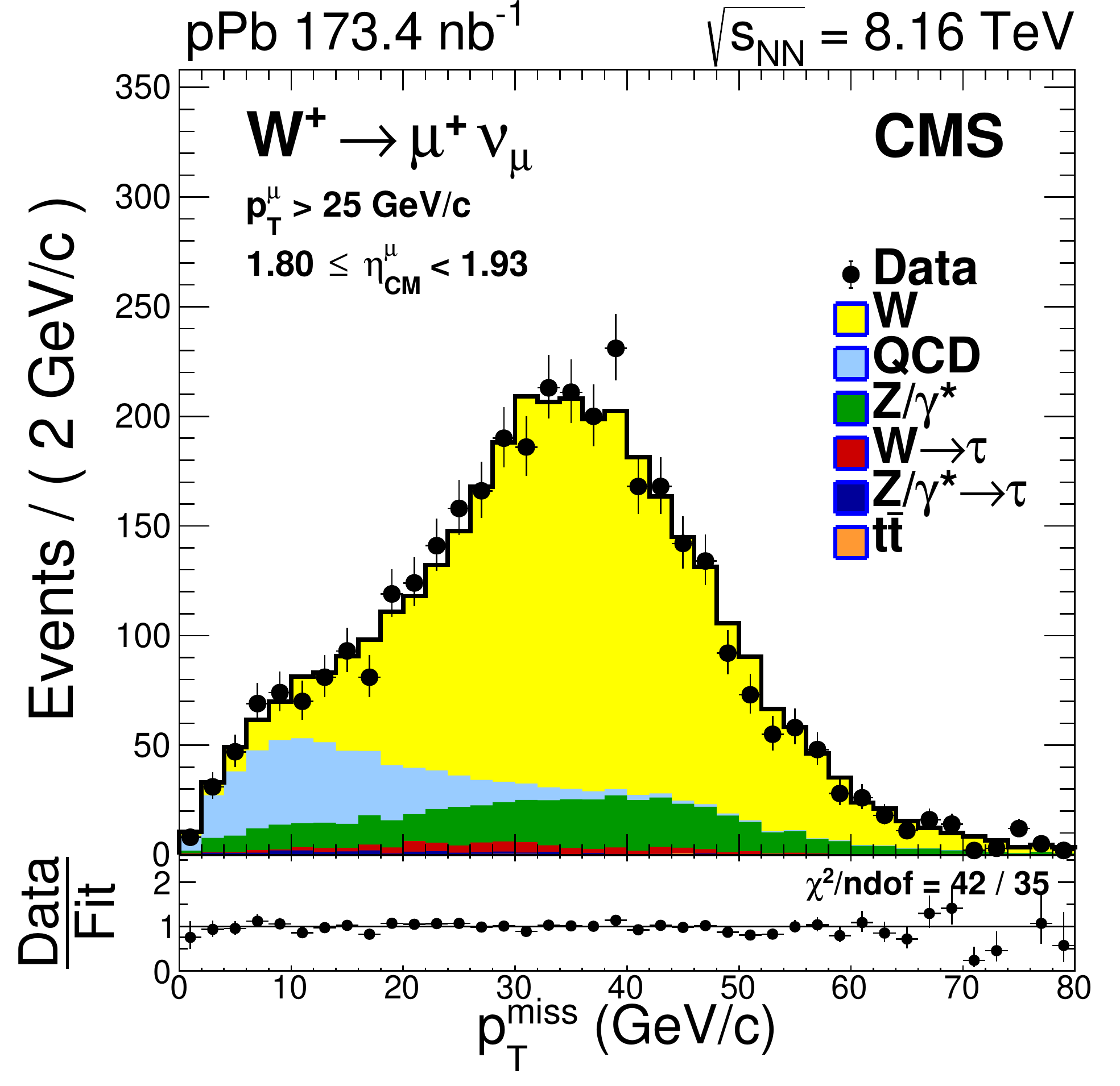}
 \caption{The missing transverse momentum \ptmiss distribution for \WToMuNuMi events within the $-0.2 < \etaMuCM < 0.0$ (\cmsLeft) range and for \WToMuNuPl events within the $1.80 < \etaMuCM < 1.93$ (\cmsRight) range. Unbinned fits to the data (black points) are performed with six contributions, stacked from bottom to top: \ttbar (orange), \DYToTauTau (dark blue), {$\Wpm\to\PGt^{\pm}\PGnGt$} (red), \DYToMuMu (green), QCD multijet (light blue) and \WToMuNupm (yellow). The \etaMuCM regions are defined such that the proton is moving towards positive pseudorapidity. Error bars represent statistical uncertainties. The lower panels display the data divided by the result of the fit. }
 \label{fig:MET}
\end{figure}

The number of \WToMuNupm events is extracted by performing an unbinned maximum likelihood fit of the observed \ptmiss distribution in each muon \etaCM bin. The total fit model includes six contributions: the signal \WToMuNupm template, the EW background \DYToMuMu, \WToTauNu and \DYToTauTau templates, the \ttbar background template, and the QCD background functional form derived from control data samples. When fitting the data, the QCD shape parameters ($\sigma_{i}$) are fixed to the extrapolated values, while the ratio of the EW  and \ttbar background yields to the signal yield is fixed to the results from simulation. Only two parameters are left free in the fit, the {\PW} boson signal and QCD background normalisations. The observed numbers of muons coming from {\PW} boson decays over the entire \etaMuCM range are: $97\,971\pm332$~{\PGmp} and $81\,147\pm301$~{\PGmm}, where the uncertainty is statistical.  Examples of the resulting \ptmiss distributions in the midrapidity ($-0.2 < \etaMuCM < 0.0$) and forward ($1.80 < \etaMuCM < 1.93$) bins, are shown in Fig.~\ref{fig:MET}, after applying all analysis corrections and selection criteria. The fit model is found to give a good description of the data, with $\chi^2$/dof values close to one.

The simulated sample of \WToMuNupm embedded into \EPOSLHC is used to derive the efficiency of the muon trigger, isolation, reconstruction, and selection criteria, as a function of \etaMuCM. These single-muon efficiencies are also directly estimated from \pPb data in a \ZToMuMu sample using the \textit{tag-and-probe} (TnP) technique, as described in Ref.~\cite{Khachatryan:2010xn}. The data and MC reconstruction efficiencies are observed to be consistent with each other, whereas the trigger efficiency is lower in the {\PZ} boson simulation by 5\% than in data at $\abs{\etaMuLab} = 1.4$. The muon isolation selection is found to reject fewer muons in the simulation, because of the smaller \pPb UE activity compared to data. In order to correct for the differences between data and simulation, the muon efficiency computed from the \WToMuNupm MC sample is multiplied by the TnP correction factors event-by-event. These correction factors are computed, in bins of muon \pt and \etaLab, from the ratio of the muon efficiencies measured in data to those calculated from simulations. The TnP scale factors produce changes in the muon efficiency ranging from $-3\%$ in the mid-rapidity region ($\abs{\etaMuLab} < 1.0$) to $+5\%$ at $\abs{\etaMuLab} = 1.4$. The TnP-corrected efficiencies vary with \etaMuCM, from $(81\pm1)\%$ to $(92\pm2)\%$.

\subsection{Systematic uncertainties}

The leading source of systematic uncertainty originates from the TnP efficiency corrections. The uncertainties on the TnP corrections are determined by propagating the uncertainties on the muon efficiencies extracted from data and simulation, derived from the fits to the invariant mass of \ZToMuMu candidates.  These uncertainties include a statistical component due to the finite size of the data sample available, as well as a systematic component estimated from variations in the fitting procedure (different signal and background functions, and different mass range used for fitting). Additionally, uncertainties are included to account for: (1) possible differences in the reconstruction of muon tracks (0.6\%), and (2) the impact of pileup and UE activity (0.3\%). Another important source of systematic uncertainty arises from the modelling of the QCD multijet \ptmiss distribution in the signal region, which is estimated by allowing the QCD shape parameters to vary within the root-mean-square of the extrapolated values in bins of \etaMuCM, and by changing the \ptmiss model to that used in Ref.~\cite{Khachatryan:2015hha}.

The uncertainty in the normalisation of the EW background is estimated from the nPDF uncertainty in the {\PZ} over {\PW} boson inclusive cross sections using the CT14 proton PDF and the EPPS16 nPDF for the lead ions, the uncertainty in the {\PW} and {\PZ} boson branching fractions to leptons~\cite{PDG2018}, and the experimental uncertainty in the \ttbar cross section in \pPb events~\cite{Sirunyan:2017xku}. The uncertainty in the vector boson \pt reweighing is derived from the difference of the results obtained applying and not applying the boson \pt correction. The uncertainty in the binning of the \ptmiss MC templates is estimated by using a \ptmiss bin size of 1\GeVc. The impact of EW corrections in \POWHEG is estimated from the difference in the efficiency when computed using \POWHEG without EW corrections~\cite{Alioli:2008gx}. The uncertainty in the \ptmiss recoil correction is determined by changing the model used to fit the hadronic recoil distribution and the profile of the recoil mean and resolution as a function of $\pt^{\PZ}$. Finally, the mismodelling of the UE activity in the simulation is estimated by reweighing the distribution of the track multiplicity instead of the energy deposited in the HF calorimeters. The integrated luminosity measurement uncertainty (3.5\%)~\cite{CMS-PAS-LUM-17-002} only affects the {\PW} boson differential cross sections and cancels out in the asymmetry measurements. The maximum relative uncertainty of the differential cross sections and absolute uncertainties on the asymmetries are presented for each source of systematic uncertainty in Table~\ref{tab:Systematics}.

\begin{table*}[htb]
\centering
\topcaption{Maximum uncertainty in the measured observables among the \etaMuCM bins determined for each source. The uncertainties in the cross sections are relative, whereas those for the asymmetries are absolute. The global integrated luminosity uncertainty of $\pm$3.5\% is not included in the total systematic uncertainty in the cross sections.}
\cmsTable{
\begin{tabular}{l c c c c c c}
\hline
Source & {\PWm} \dd{\sigma}{\etaCM} [\%] & {\PWp} \dd{\sigma}{\etaCM} [\%] & {\PWm} \RFB & {\PWp} \RFB & {\PW} \RFB & $\frac{ N_{\mu}^{+} - N_{\mu}^{-}}{N_{\mu}^{+} + N_{\mu}^{-}}$\\
\hline
Boson \pt reweighing &  0.5  &  0.4  &  0.001  &  0.001  &  0.001  &  0.001 \\
EW background &  0.4  &  0.3  &  0.002  &  0.001  &  0.001  &  0.000 \\
POWHEG EW correction &  0.9  &  0.5  &  0.007  &  0.004  &  0.006  &  0.003 \\
Efficiency &  3.0  &  3.2  &  0.026  &  0.037  &  0.030  &  0.011 \\
Event activity reweighing &  0.6  &  0.4  &  0.002  &  0.002  &  0.001  &  0.002 \\
\ptmiss template binning &  0.1  &  0.1  &  0.002  &  0.001  &  0.001  &  0.001 \\
QCD background &  1.2  &  0.7  &  0.016  &  0.007  &  0.009  &  0.006 \\
Hadronic recoil correction &  0.2  &  0.3  &  0.002  &  0.004  &  0.002  &  0.002 \\
Total systematic uncertainty &  3.3  &  3.3  &  0.030  &  0.038  &  0.031  &  0.013 \\
Statistical uncertainty &  2.4  &  2.0  &  0.026  &  0.029  &  0.019  &  0.015 \\
\hline
\end{tabular}
}
\label{tab:Systematics}
\end{table*}

\section{Results}

The {\Wpm} boson differential production cross sections are computed as functions of \etaMuCM. The differential \WToMuNupm cross sections are determined from
\begin{equation}
\dd{\sigma^{\WToMuNupm}}{\etaMuCM}\left(\etaMuCM\right) = \frac{N_{\PGm}\left(\etaMuCM\right)}{\Lumi \Delta\etaMuCM},
\end{equation}
where $N_{\PGm}\left(\etaMuCM\right)$ is the efficiency-corrected muon yield in bins of \etaMuCM, \Lumi is the recorded integrated luminosity, and $\Delta\etaMuCM$ is the width of the measured bin.

\begin{figure}[htbp]
 \centering
 \includegraphics[width=1.2\cmsFigWidth]{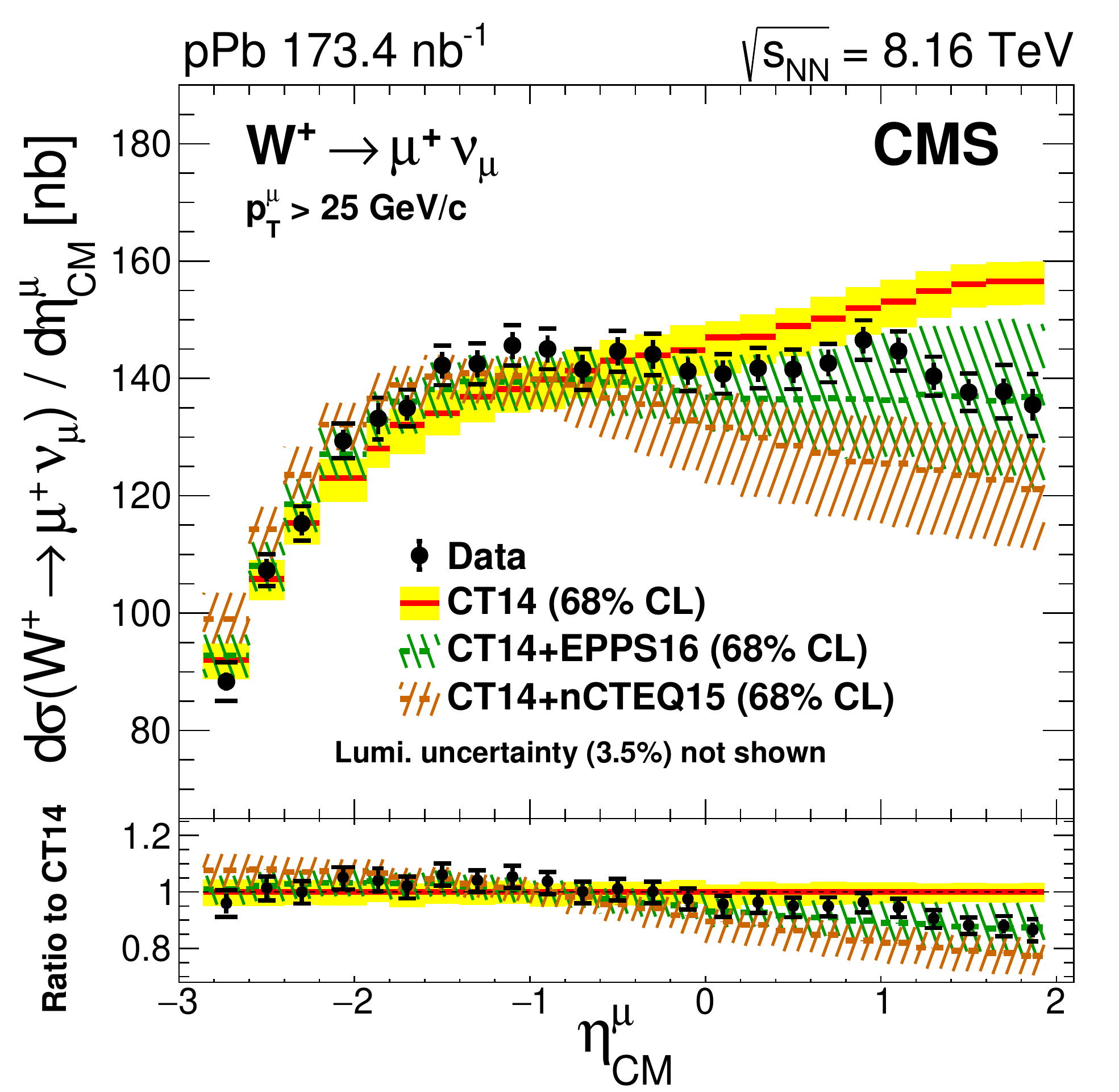}
 \includegraphics[width=1.2\cmsFigWidth]{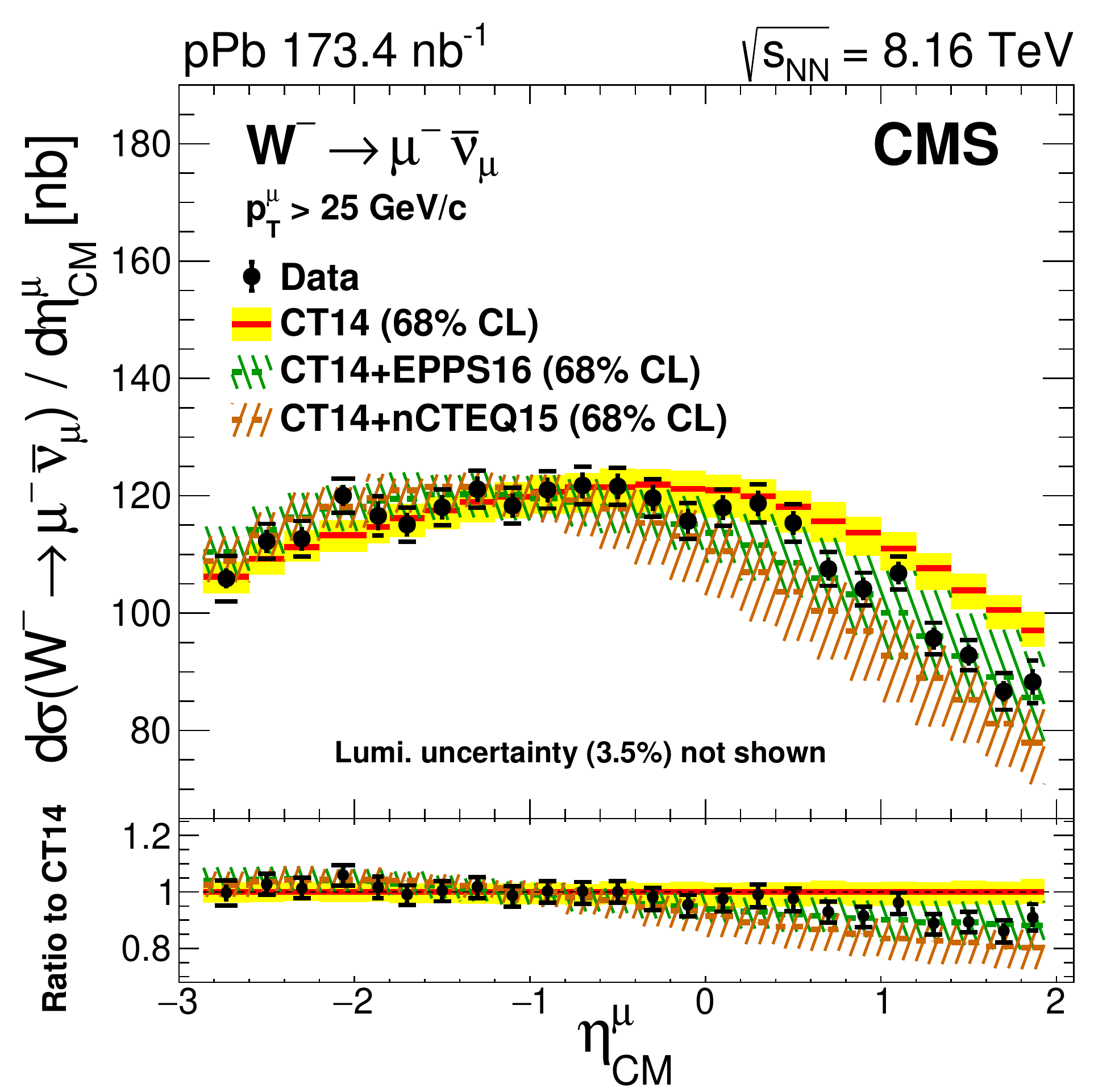}
 \caption{Differential production cross sections for \WToMuNuPl (\cmsLeft) and \WToMuNuMi (\cmsRight), as a function of the muon pseudorapidity in the centre-of-mass frame. The small horizontal lines represent the statistical and systematic uncertainties summed in quadrature, whereas the error bars show the statistical uncertainties only. The global integrated luminosity uncertainty of $\pm$3.5\% is not shown. The NLO calculations with CT14 PDF, and CT14+EPPS16 and CT14+nCTEQ15 nPDFs, are also displayed, including their 68\% confidence interval PDF uncertainty bands. The bottom panels show the ratio of data, CT14+EPPS16 and CT14+nCTEQ15 with respect to CT14.}
 \label{fig:Cross_Section_WithTheory}
\end{figure}

The cross sections for the \WToMuNu decays for {\PWp} and {\PWm} bosons are compared in Fig.~\ref{fig:Cross_Section_WithTheory} with NLO perturbative QCD predictions calculated with the MC program \MCFM v8.0~\cite{Boughezal:2016wmq} using the CT14~\cite{Dulat:2015mca} proton PDF. Also shown are two calculations that include nuclear modifications in the PDF, based on the nCTEQ15~\cite{Kovarik:2015cma} and  EPPS16~\cite{Eskola:2016oht} nPDF sets (labelled as CT14+nCTEQ15 and CT14+EPPS16, respectively). Both EPPS16 and nCTEQ15 are Hessian NLO nPDF sets, but the former includes more measurements in the fit (containing LHC EW boson~\cite{Khachatryan:2015hha,Khachatryan:2015pzs,Aad:2015gta} and dijet~\cite{Chatrchyan:2014hqa} data), as well as more free parameters (20 for EPPS16, 17 for nCTEQ15). In addition, nuclear modifications of valence and sea quarks are allowed to be different in EPPS16 for up and down quarks, while nCTEQ15 assumes flavour independence for the sea quarks. The nPDF uncertainties are propagated using the PDF4LHC recommendations for Hessian nPDF sets as prescribed in Ref.~\cite{Butterworth:2015oua}. As can be seen in Fig.~\ref{fig:Cross_Section_WithTheory}, the predicted CT14+nCTEQ15 and CT14+EPPS16 cross sections are systematically below the calculation using CT14 PDF at large positive muon rapidities because of the depletion of the antiquark PDF in nuclei at small $x = (M_{\PW}/\sqrtsNN)\exp(-y_{\PW}) \simeq (M_{\PW}/\sqrtsNN)\exp(-\etaMuCM) \approx 10^{-3}$. Conversely, the predicted cross sections from calculations including nPDF modifications are above those using CT14 PDF in the negative rapidity region, because of the slight quark antishadowing at large $x \approx 0.1$. When compared to data, all theoretical calculations reproduce the measurement at backward rapidity within uncertainties, while at forward rapidity the calculations including nPDF effects appear to be favoured.

\begin{figure*}[htbp]
 \centering
 \includegraphics[width=0.32\textwidth]{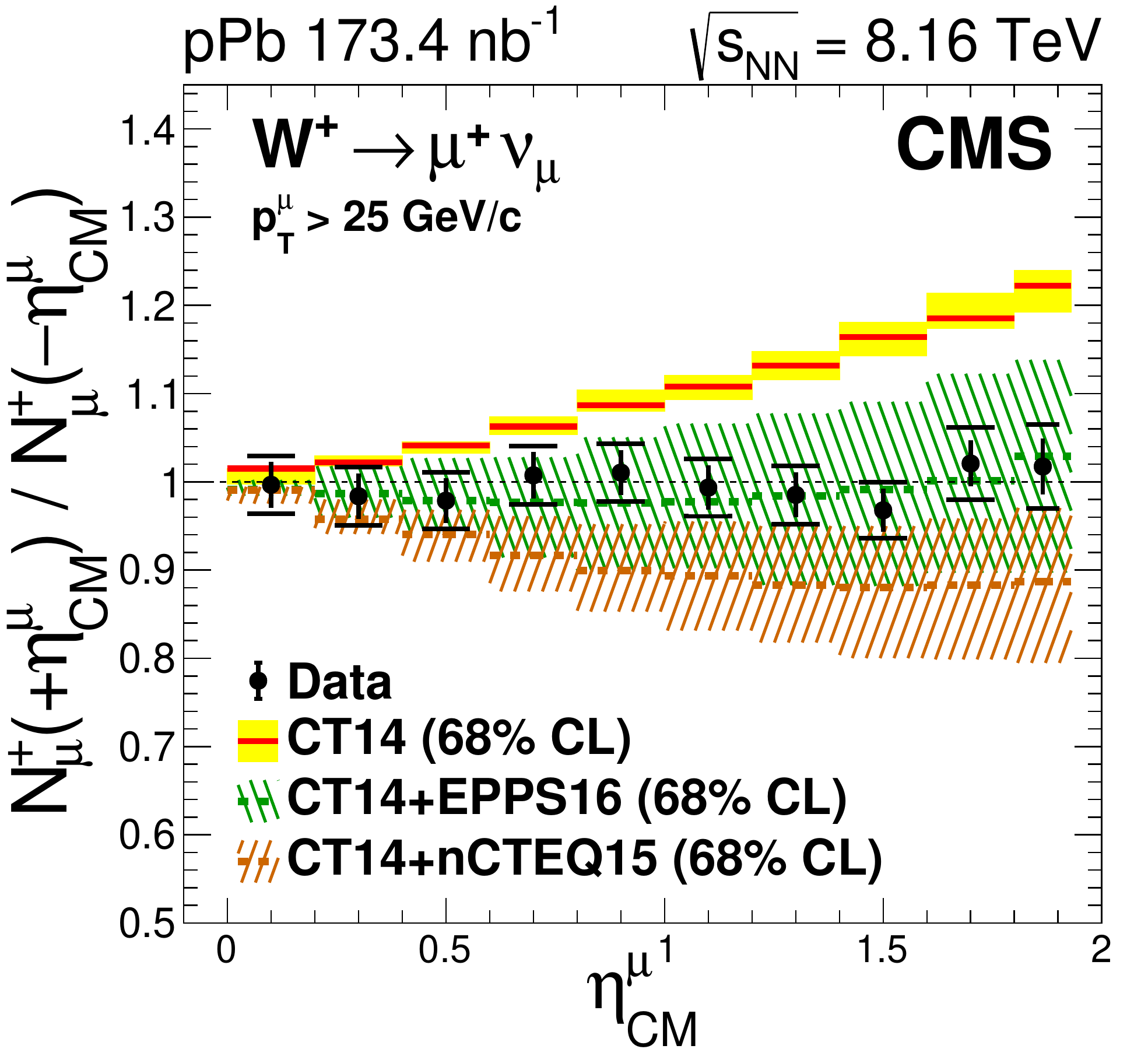}
 \includegraphics[width=0.32\textwidth]{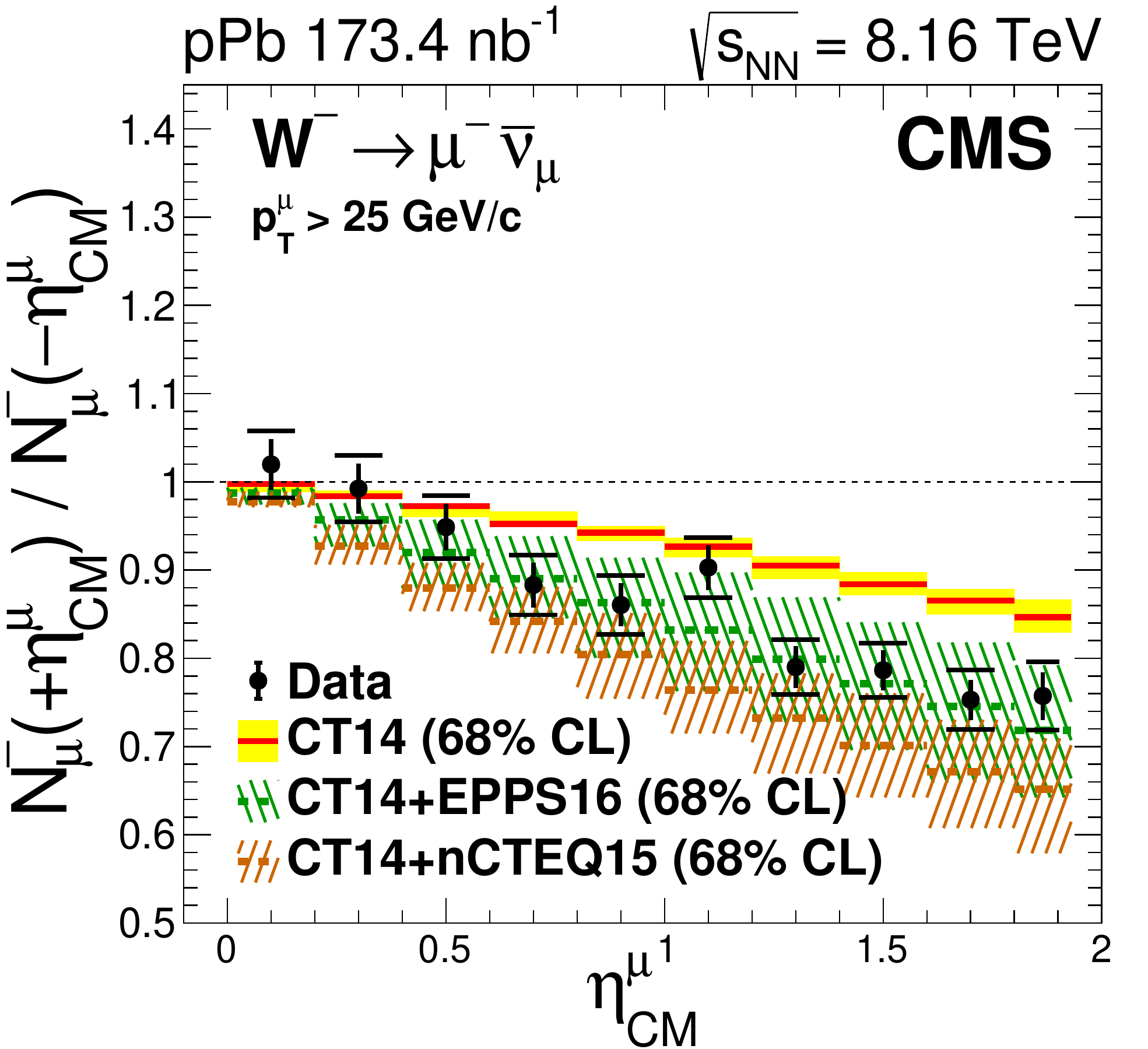}
  \includegraphics[width=0.32\textwidth]{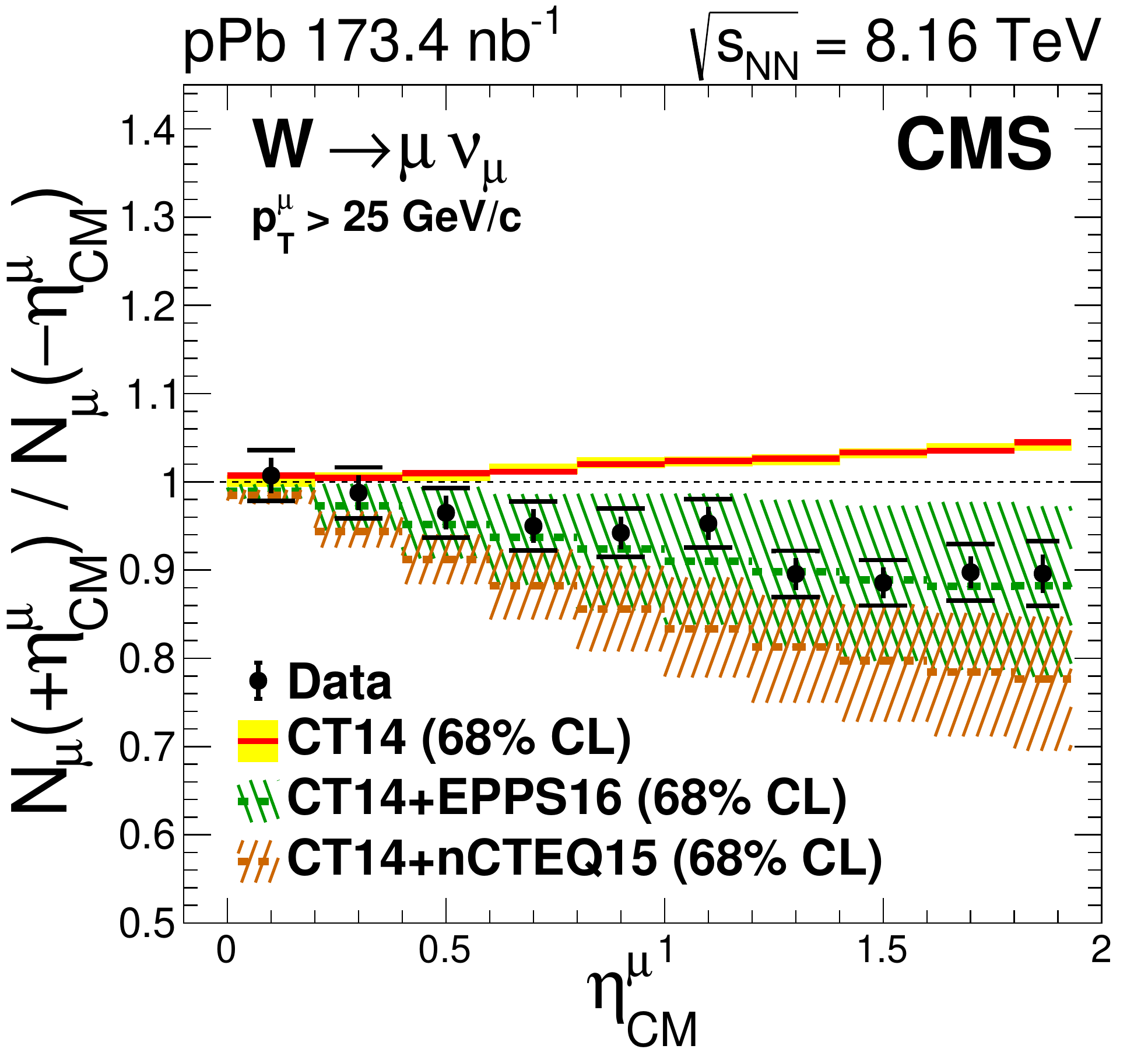}
 \caption{Forward-backward ratios, \RFBChgMu, for the positively (left) and negatively (middle) charged muons, and the forward-backward ratio for muons of both signs, \RFBAllMu (right), as a function of \etaMuCM. The small horizontal lines represent the statistical and systematic uncertainties summed in quadrature, whereas the error bars show the statistical uncertainties only. The NLO calculations with CT14 PDF, CT14+EPPS16 nPDF, and CT14+nCTEQ15 nPDF, are also displayed, including their 68\% confidence interval PDF uncertainty bands.}
 \label{fig:ForwardBackward_Ratio}
\end{figure*}

The muon forward-backward ratios, defined as \RFBChgMu for both positive and negative muons, are compared in the upper panel of Fig.~\ref{fig:ForwardBackward_Ratio} to the CT14 PDF, and CT14+EPPS16 and CT14+nCTEQ15 nPDF calculations. These observables probe the ratio of the nuclear modifications of the quark PDFs in the Pb nucleus from small to large $x$ values. The results for muons of both charges favour the predictions including nuclear modifications over the free-proton PDF calculations. Based on the precision of the experimental results, the measurements provide constraints on both the CT14+EPPS16 and CT14+nCTEQ15 nPDF sets, especially in the proton-going region (small $x$).

The yields of positively and negatively charged muons are further combined to measure the forward-backward ratio for all muons \RFBAllMu. This observable has a couple of advantages compared to \RFBChgMu: it is less sensitive to the quark content in the proton and nuclei, and it has better statistical precision. The results for this asymmetry are presented in the right panel of Fig.~\ref{fig:ForwardBackward_Ratio}, and they strongly deviate from the CT14 PDF predictions, favouring the CT14+nCTEQ15 and CT14+EPPS16 nPDF sets. Moreover, the experimental uncertainties are significantly smaller than the theoretical nPDF uncertainties. Consequently, these measurements could constrain the quark and antiquark distributions in nuclei, and will be valuable inputs for global fits to the data.

\begin{figure}[htb]
 \centering
 \includegraphics[width=1.2\cmsFigWidth]{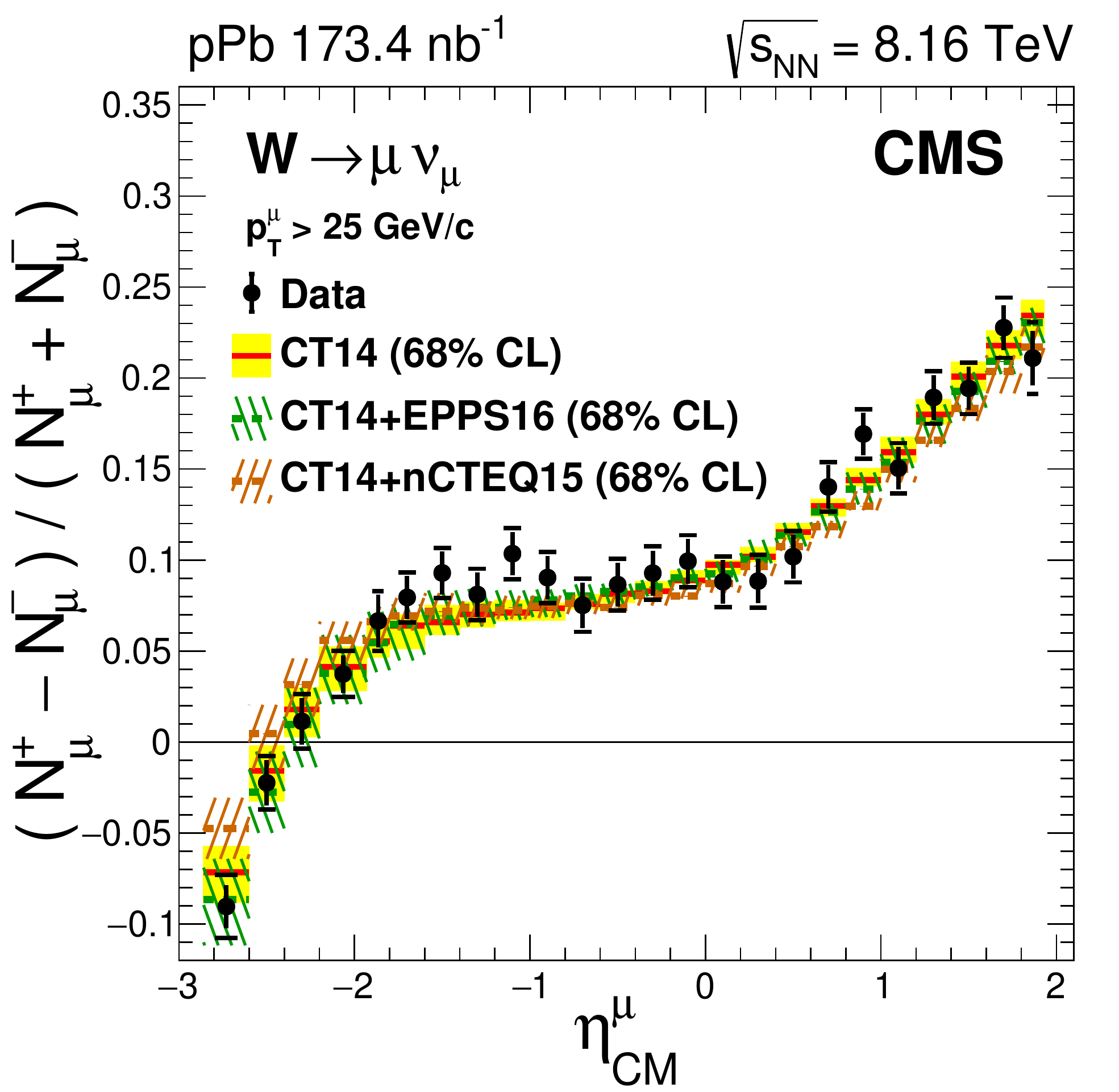}
 \caption{Muon charge asymmetry, \ChgAsym, as a function of the muon pseudorapidity in the centre-of-mass frame. The small horizontal lines represent the statistical and systematic uncertainties summed in quadrature, whereas the error bars show the statistical uncertainties only. The NLO calculations with CT14 PDF, CT14+EPPS16 nPDF, and CT14+nCTEQ15 nPDF, are also displayed, including their 68\% confidence interval PDF uncertainty bands.}
 \label{fig:ChargeAsymmetry}
\end{figure}

The muon charge asymmetry, defined as $\mathcal{A}\equiv\ChgAsym$, reflects the differences in the production of {\PWp} and {\PWm} bosons. Figure~\ref{fig:ChargeAsymmetry} shows the measurement of the muon charge asymmetry as a function of \etaMuCM compared to the \MCFM~\cite{Boughezal:2016wmq} predictions calculated using CT14 PDF alone and including nuclear modifications described by the EPPS16 and nCTEQ15 nPDFs. All calculations reproduce the present measurements within uncertainties in the entire muon $\eta$ range, including when the CT14 proton PDF set is used, because nuclear modifications of the PDFs mostly cancel in this quantity.

\begin{figure}[htb]
 \centering
 \includegraphics[width=1.2\cmsFigWidth]{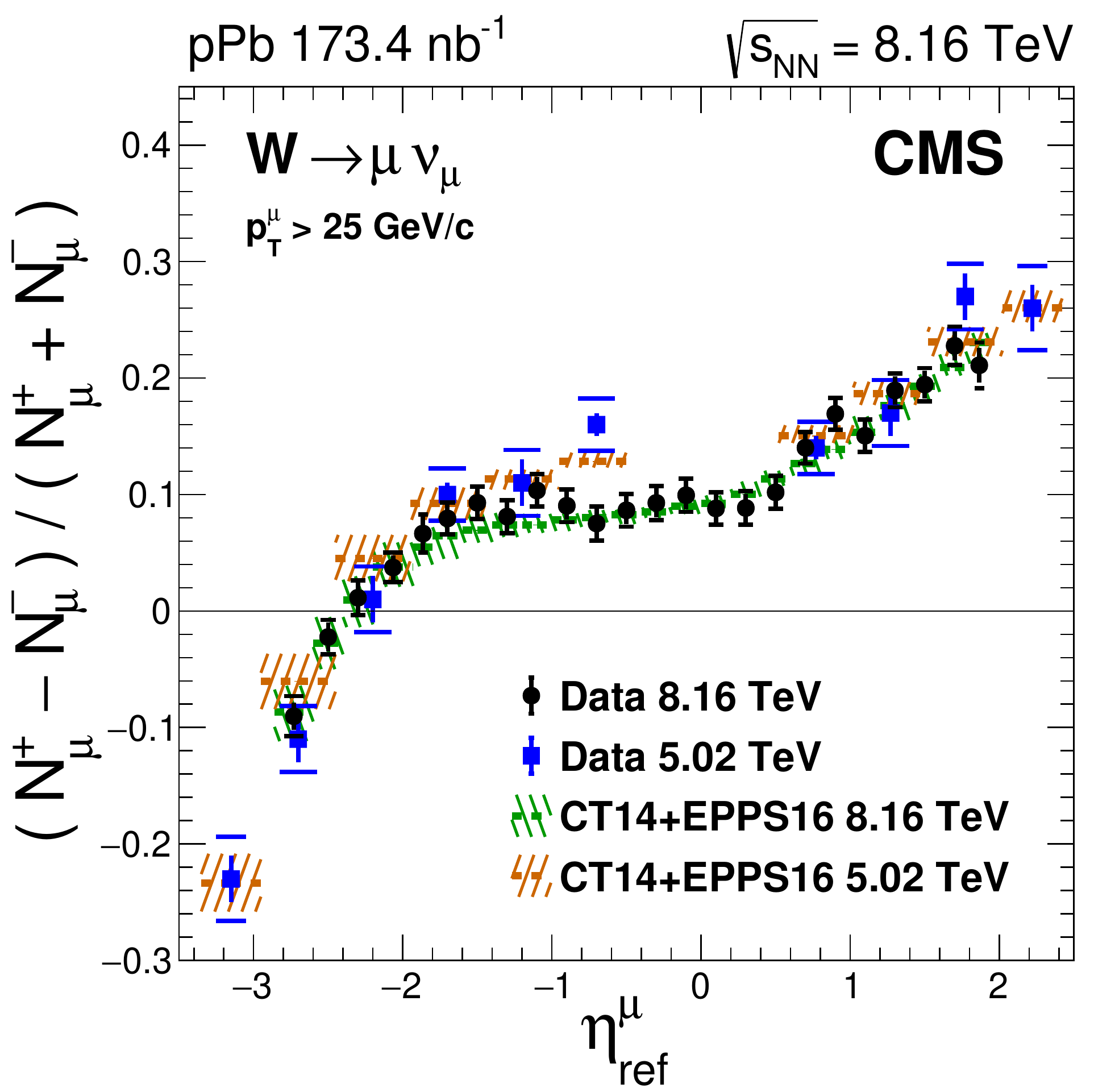}
 \caption{Comparison of the muon charge asymmetry measured at $\sqrtsNN = 8.16\TeV$ (circles) and at $\sqrtsNN = 5.02\TeV$~\cite{Khachatryan:2015hha} (squares). The muon pseudorapidity of the measurements at 5.02\TeV has been shifted (see text for details)~\cite{Arleo:2015dba}. The small horizontal lines represent the statistical and systematic uncertainties summed in quadrature, whereas the error bars show the statistical uncertainties only. The NLO calculations with CT14+EPPS16 nPDF at 8.16\TeV and at 5.02\TeV, are also displayed, including their 68\% confidence interval PDF uncertainty bands.}
 \label{fig:ChargeAsymmetry_With5TeV}
\end{figure}

The tension between data and theoretical calculations reported at negative muon $\eta$ in \pPb collisions at $\sqrtsNN = 5.02\TeV$~\cite{Khachatryan:2015hha} is not observed in the present measurements. The present use of the CT14 proton PDF set decreases the value of the charge asymmetry compared to the predictions based on CT10 in Ref.~\cite{Khachatryan:2015hha}. Moreover, the theoretical uncertainties are also enlarged in the EPPS16 nPDF sets and the theoretical calculations using the CT14+EPPS16 nPDF sets agree better with the measurements at $\sqrtsNN = 5.02\TeV$, as compared to the EPS09 nPDF sets used in the analysis at $\sqrtsNN = 5.02\TeV$. It has been shown in Ref.~\cite{Arleo:2015dba} that the measurements of the lepton charge asymmetry at different collision energies ($\sqrt{s^\prime}$) are simply related by a shift in the lepton pseudorapidity, $\mathcal{A}(\eta_\Pl, \sqrt{s^\prime}) = \mathcal{A}(\eta_{\text{ref}}, \sqrt{s})$, where $\eta_{\text{ref}} = \eta_\Pl + \ln(\sqrt{s}/\sqrt{s^\prime})$ if $\eta_\Pl > 0$ and $\eta_{\text{ref}} = \eta_\Pl - \ln(\sqrt{s}/\sqrt{s^\prime})$ if $\eta_\Pl < 0$. The result of this shift is shown in Fig.~\ref{fig:ChargeAsymmetry_With5TeV}, demonstrating that the present results and the measurements performed at $\sqrtsNN = 5.02\TeV$~\cite{Khachatryan:2015hha} obey this scaling property.

The agreement between data and theoretical calculations is quantified through a $\chi^{2}$ test performed for each observable taking into account both experimental (including luminosity) and theoretical uncertainties and their bin-to-bin correlations, obtained following the prescription for Hessian PDF sets~\cite{Buckley:2014ana} and rescaled to 68\% confidence intervals. The results of the $\chi^{2}$ test and the dof of each observable are shown in Table~\ref{tab:modelComparison_CT14}. The CT14+EPPS16 and CT14+nCTEQ15 nPDF predictions prove compatible with the data, while the CT14 PDF calculations do not describe the measurements well. These experimental results thus provide for the first time clear evidence of the nuclear modification of quark PDFs from the measurements of EW boson production in nuclear collisions. Bin-to-bin correlations have been found to have a large impact on the obtained $\chi^{2}$ values, especially from nPDF uncertainties in the NLO calculations, which are strongly correlated inside each of the shadowing (positive \etaMuCM) and antishadowing (negative \etaMuCM) regions, and anticorrelated between these two regions.

\begin{table*}[htb]
\topcaption{Results of the $\chi^{2}$ statistical test between the measurements and the nPDF calculations from the CT14 PDF, CT14+EPPS16 nPDF, and CT14+nCTEQ15 nPDF sets. The value of the $\chi^{2}$, the number of degrees of freedom (dof) and the $\chi^{2}$ probability (Prob.), are presented for the {\Wpm} boson differential cross sections, the muon charge asymmetries, the charged muon forward-backward ratios, and the forward-backward ratios of all muons, respectively.}
\cmsTable{
\centering
\begin{tabular}{l *3c *3c *3c}
\hline
\multirow{2}{*}{Observable}  & \multicolumn{3}{c}{CT14} & \multicolumn{3}{c}{CT14+EPPS16} & \multicolumn{3}{c}{CT14+nCTEQ15} \\
                                             & $\chi^{2}$ & dof & Prob. [\%] & $\chi^{2}$ & dof & Prob. [\%] & $\chi^{2}$ & dof & Prob. [\%] \\
\hline
\ddinline{\sigma^{\WToMuNupm}(\etaMuCM)}{\etaMuCM} & 135 & 48 & 3\ten{-8} & 32 & 48 & 96 & 40 & 48 & 79 \\
\ChgAsym & 23 & 24 & 54 & 18 & 24 & 80 & 29 & 24 & 23  \\
\RFBChgMu & 98 & 20 & 3\ten{-10} & 11 & 20 & 95 & 14 & 20 & 83  \\
\RFBAllMu & 87 & 10 & 2\ten{-12} & 3 & 10 & 99 & 5 & 10 & 90  \\
\hline
\end{tabular}
}
\label{tab:modelComparison_CT14}
\end{table*}

\begin{figure}[htb]
 \centering
 \includegraphics[width=1.2\cmsFigWidth]{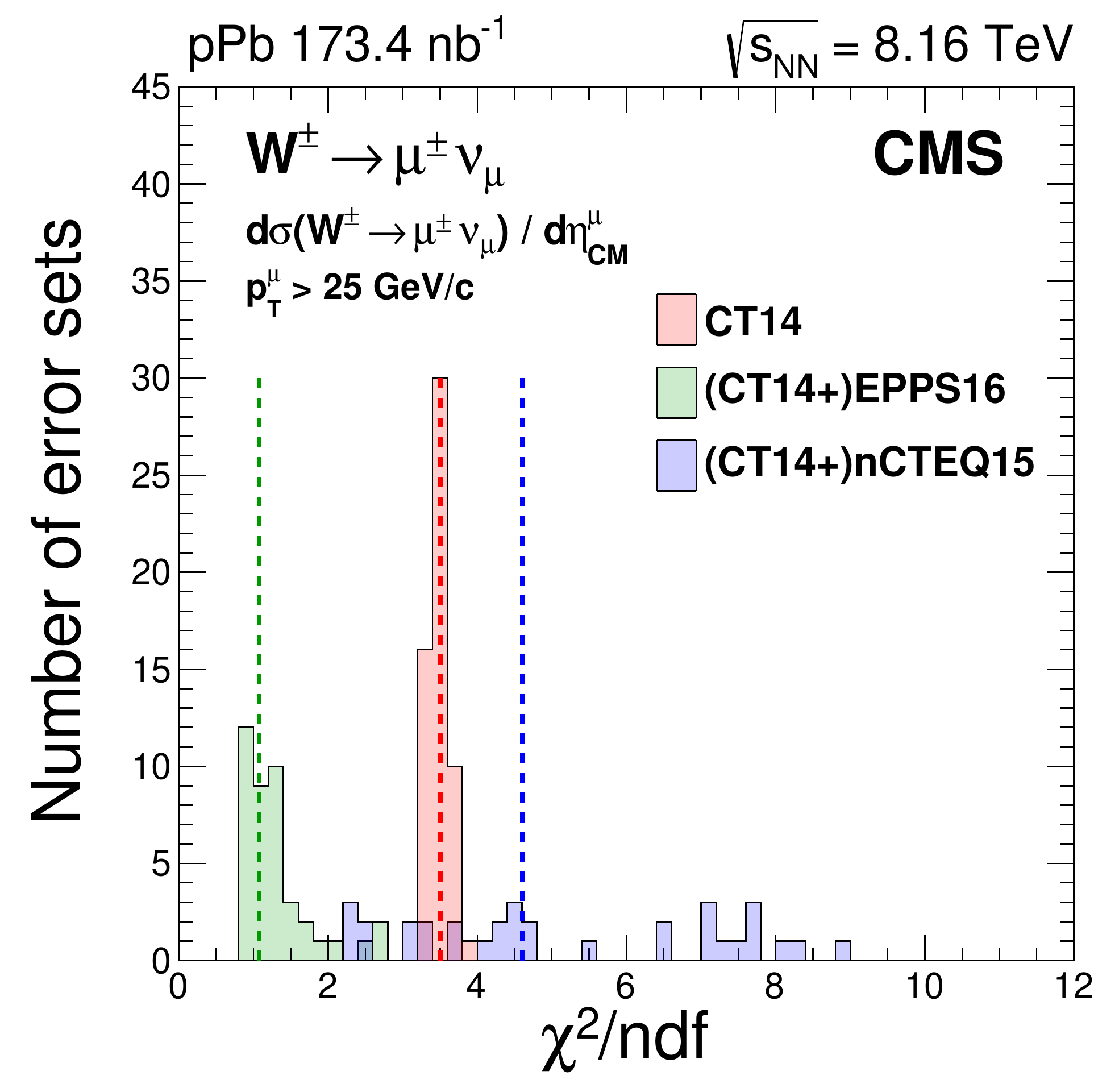}
 \caption{Distribution of the $\chi^2$/dof values from the comparison of data (cross section measurements) and theoretical calculations, for the CT14, nCTEQ15, and EPPS16 individual error sets. The vertical dashed lines represent the prediction corresponding to the central set of CT14, nCTEQ15, and EPPS16.}
 \label{fig:chi2histo}
\end{figure}

Furthermore, the possible sources of differences between data and the (n)PDFs are investigated. In the Hessian representation, a central PDF is given along with error sets, each of which corresponds to an eigenvector of the covariance matrix in parameter space~\cite{Pumplin:2001ct}. The values of $\chi^{2}$/dof corresponding to the compatibility between the cross section measurements and the calculations using each of the individual sets of CT14, nCTEQ15, and EPPS16 (57, 33 and 41 error sets, respectively) have been determined. Figure~\ref{fig:chi2histo} shows the distribution of the $\chi^{2}$/dof values for the central and error sets. The $\chi^{2}$/dof values obtained are for individual sets, thus ignoring theoretical uncertainties and their correlations. While most of the EPPS16 individual sets lead to a good agreement with data (with $\chi^2$/dof around unity), only those nCTEQ15 sets that exhibit the smaller quark shadowing at small $x$ are more compatible with the data, yet with $\chi^2/\text{dof} \gtrsim 2$. All CT14 PDF sets lead to a narrow distribution centred around $\chi^2/\text{dof} \simeq3$, because of the strong constraints imposed by the large experimental data sets used to extract them. The current measurements of {\Wpm} boson production in \pPb collisions will permit further constraints on the quark and antiquark nPDFs and the amount of quark shadowing in the nuclei.

\section{Summary}

A study of {\Wpm} boson production in \pPb collisions at a nucleon-nucleon centre-of-mass energy of $\sqrtsNN = 8.16\TeV$ is reported, using the muon decay channel for muons with transverse momenta greater than 25\GeVc and for absolute values of the pseudorapidity in the laboratory frame $\abs{\etaMuLab} < 2.4$. The differential production cross sections for positively and negatively charged \WToMuNu decays, the muon charge asymmetry, and the muon forward-backward ratios, are measured as functions of the muon pseudorapidity in the centre-of-mass frame, in the range $-2.86 < \etaMuCM < 1.93$.

The measurements are compared to theoretical predictions from both proton parton distribution functions (PDFs) (CT14) and nuclear PDF (CT14+EPPS16 , CT14+nCTEQ15) sets. The cross sections and the forward-backward asymmetries exhibit significant deviations from the CT14 prediction, revealing nuclear modifications of the PDFs unambiguously for the first time in the production of electroweak bosons in nuclear collisions. Both the CT14+EPPS16, and the CT14+nCTEQ15 calculations show a good overall agreement with the data, with the measurements favouring the former nPDF set. In the latter case, only the individual sets that exhibit the smallest nuclear PDF modifications at small values of $x$ (in the shadowing region) turn out to be compatible with experimental measurements. The small experimental uncertainties allow for a significant reduction in the current uncertainties on the quark and antiquark nuclear PDFs in the range $10^{-3} \lesssim x \lesssim 10^{-1}$.

\begin{acknowledgments}

We congratulate our colleagues in the CERN accelerator departments for the excellent performance of the LHC and thank the technical and administrative staffs at CERN and at other CMS institutes for their contributions to the success of the CMS effort. In addition, we gratefully acknowledge the computing centres and personnel of the Worldwide LHC Computing Grid for delivering so effectively the computing infrastructure essential to our analyses. Finally, we acknowledge the enduring support for the construction and operation of the LHC and the CMS detector provided by the following funding agencies: BMBWF and FWF (Austria); FNRS and FWO (Belgium); CNPq, CAPES, FAPERJ, FAPERGS, and FAPESP (Brazil); MES (Bulgaria); CERN; CAS, MoST, and NSFC (China); COLCIENCIAS (Colombia); MSES and CSF (Croatia); RPF (Cyprus); SENESCYT (Ecuador); MoER, ERC IUT, PUT and ERDF (Estonia); Academy of Finland, MEC, and HIP (Finland); CEA and CNRS/IN2P3 (France); BMBF, DFG, and HGF (Germany); GSRT (Greece); NKFIA (Hungary); DAE and DST (India); IPM (Iran); SFI (Ireland); INFN (Italy); MSIP and NRF (Republic of Korea); MES (Latvia); LAS (Lithuania); MOE and UM (Malaysia); BUAP, CINVESTAV, CONACYT, LNS, SEP, and UASLP-FAI (Mexico); MOS (Montenegro); MBIE (New Zealand); PAEC (Pakistan); MSHE and NSC (Poland); FCT (Portugal); JINR (Dubna); MON, RosAtom, RAS, RFBR, and NRC KI (Russia); MESTD (Serbia); SEIDI, CPAN, PCTI, and FEDER (Spain); MOSTR (Sri Lanka); Swiss Funding Agencies (Switzerland); MST (Taipei); ThEPCenter, IPST, STAR, and NSTDA (Thailand); TUBITAK and TAEK (Turkey); NASU and SFFR (Ukraine); STFC (United Kingdom); DOE and NSF (USA).

\hyphenation{Rachada-pisek} Individuals have received support from the Marie-Curie programme and the European Research Council and Horizon 2020 Grant, contract Nos.\ 675440 and 765710 (European Union); the Leventis Foundation; the A.P.\ Sloan Foundation; the Alexander von Humboldt Foundation; the Belgian Federal Science Policy Office; the Fonds pour la Formation \`a la Recherche dans l'Industrie et dans l'Agriculture (FRIA-Belgium); the Agentschap voor Innovatie door Wetenschap en Technologie (IWT-Belgium); the F.R.S.-FNRS and FWO (Belgium) under the ``Excellence of Science -- EOS" -- be.h project n.\ 30820817; the Beijing Municipal Science \& Technology Commission, No. Z181100004218003; the Ministry of Education, Youth and Sports (MEYS) of the Czech Republic; the Lend\"ulet (``Momentum") Programme and the J\'anos Bolyai Research Scholarship of the Hungarian Academy of Sciences, the New National Excellence Program \'UNKP, the NKFIA research grants 123842, 123959, 124845, 124850, 125105, 128713, 128786, and 129058 (Hungary); the Council of Science and Industrial Research, India; the HOMING PLUS programme of the Foundation for Polish Science, cofinanced from European Union, Regional Development Fund, the Mobility Plus programme of the Ministry of Science and Higher Education, the National Science Center (Poland), contracts Harmonia 2014/14/M/ST2/00428, Opus 2014/13/B/ST2/02543, 2014/15/B/ST2/03998, and 2015/19/B/ST2/02861, Sonata-bis 2012/07/E/ST2/01406; the National Priorities Research Program by Qatar National Research Fund; the Programa Estatal de Fomento de la Investigaci{\'o}n Cient{\'i}fica y T{\'e}cnica de Excelencia Mar\'{\i}a de Maeztu, grant MDM-2015-0509 and the Programa Severo Ochoa del Principado de Asturias; the Thalis and Aristeia programmes cofinanced by EU-ESF and the Greek NSRF; the Rachadapisek Sompot Fund for Postdoctoral Fellowship, Chulalongkorn University and the Chulalongkorn Academic into Its 2nd Century Project Advancement Project (Thailand); the Welch Foundation, contract C-1845; and the Weston Havens Foundation (USA).

\end{acknowledgments}

\bibliography{auto_generated} \cleardoublepage \appendix\section{The CMS Collaboration \label{app:collab}}\begin{sloppypar}\hyphenpenalty=5000\widowpenalty=500\clubpenalty=5000\vskip\cmsinstskip
\textbf{Yerevan Physics Institute, Yerevan, Armenia}\\*[0pt]
A.M.~Sirunyan, A.~Tumasyan
\vskip\cmsinstskip
\textbf{Institut f\"{u}r Hochenergiephysik, Wien, Austria}\\*[0pt]
W.~Adam, F.~Ambrogi, E.~Asilar, T.~Bergauer, J.~Brandstetter, M.~Dragicevic, J.~Er\"{o}, A.~Escalante~Del~Valle, M.~Flechl, R.~Fr\"{u}hwirth\cmsAuthorMark{1}, V.M.~Ghete, J.~Hrubec, M.~Jeitler\cmsAuthorMark{1}, N.~Krammer, I.~Kr\"{a}tschmer, D.~Liko, T.~Madlener, I.~Mikulec, N.~Rad, H.~Rohringer, J.~Schieck\cmsAuthorMark{1}, R.~Sch\"{o}fbeck, M.~Spanring, D.~Spitzbart, A.~Taurok, W.~Waltenberger, J.~Wittmann, C.-E.~Wulz\cmsAuthorMark{1}, M.~Zarucki
% ...
\vskip\cmsinstskip
% ...
1:  Also at Vienna University of Technology, Vienna, Austria\\
% ...
\end{sloppypar}
\end{document}